\documentclass[12pt]{iopart}
\usepackage[latin1]{inputenc}
\usepackage[T1]{fontenc}
\usepackage[english]{babel}
\begin{document}
\title[Hamiltonian study of the Generalized scalar-tensor theory...]{Hamiltonian study of the Generalized scalar-tensor theory with potential in a Bianchi type I model}
\author{Stéphane Fay}
\address{66 route de la Montée Jaune\\
37510 Savonnières
\footnote{E-mail: Steph.Fay@Wanadoo.fr}}
\begin{abstract}
We study the generalized scalar tensor theory with a potential in the Bianchi type I model by using the ADM formalism. We examine the conditions for the Universe to be in expansion, isotropic and with a positive potential at late time in the Brans-Dicke and Einstein frames. In particular, we analyse the two important cases where metric functions tend, in an asymptotic way, toward power or exponential laws in the Einstein frame.\\
\\
\\
\\
Published in Classical and Quantum Gravity copyright 2000 IOP Publishing Ltd\\
Classical and Quantum Gravity, Vol 17, 4, 891, 2000.\\
http://www.iop.org
\end{abstract}
\maketitle
\section{Introduction} \label{s0}
We study the Generalized Scalar Tensor theory with a potential depending on a scalar field in the Bianchi type I model. This theory has the same form as Brans-Dicke theory but with a coupling function depending on a scalar field. Its dynamical behaviour with matter, but without a potential, in the homogeneous Bianchi type models has been studied by Wands and Mimoso \cite{MimWan95} and in the FLRW models by Barrow and Parson \cite{BarPar97}.

The potential can be considered as an effective cosmological constant. Such a constant can rule out the Universe age problem \cite{Mof95}. The cosmological constant is a source of negative pressure able to accelerate the expansion of Universe and hence to give birth to inflation. Then, the Universe would seem younger than it is. Moreover, new observations \cite{Per99}\cite{Rie98} would show that the Universe is undergoing inflation and this the presence of a positive cosmological constant although this accelerated behaviour remains to be confirmed. However the present value on this constant is in contradiction with the value predicted by particle physics for the early Universe. This is the reason why a model with a varying effective cosmological constant is so interesting. One recalls that the empty generalized scalar tensor theory can naturally generate inflation without a potential: this is what is usually called kinetic inflation \cite{Lev95}\cite{JLev95}.

The aim of this work is to analyse under which conditions the Universe can isotropize and be in expansion with a positive potential at late time in the Einstein and Brans-Dicke frames. Once they are derived, we look for additional conditions such that the metric functions tend asymptotically toward exponential or power laws of the proper time in the Einstein frame. We discuss whether such theories can respect the solar system tests. When no matter field is present, this means that the coupling function $\omega$ becomes infinite or at least greater than 500 and $\omega_{\phi}\omega^{-3}$ tends to vanish, where $\omega_{\phi}$ is the derivative of $\omega$ with respect to the scalar field. No such conditions are known in a theory with a potential but if we add one and consider it as an effective cosmological constant, the observations show that it should be rather small at late time. So it seems reasonable to assume that these three conditions, $\omega\rightarrow \infty$, $\omega_\phi\omega^{-3}\rightarrow 0$ and a small potential at late time, are necessary but not sufficient for the solar system tests to be respected in a generalized scalar tensor theory with a potential. Note that even if a generalized scalar tensor theory tends toward a relativistic behaviour, it does not mean that its solutions, in these conditions, will tend toward relativistic one as shown in \cite{Far99}.

To obtain these results, we will employ a Hamiltonian formalism and more precisely the ADM formalism. It is often used in quantum cosmology, to find the wave-function of the Universe but less so to study classical problems such as the search for exact solutions or dynamics of the classical field equations \cite{UggJanRos95}. Usually Lagrangian methods are preferred.

This paper is organised as follows: in section \ref{s1}, we establish the field equation of the ADM formalism in the Einstein frame. In section \ref{s2} we analyse the dynamics of the theory in this frame and when it isotropizes. In section \ref{s3}, we examine which conditions have to be respected by the Hamiltonian and the scalar field so that the Universe can isotropize and be in expansion at late times in the Brans-Dicke frame with a positive potential. In section \ref{s4}, we discuss the best conditions in each frame so that the Universe can be isotropic, expanding, and with positive potential at late times and say a few words about the production of exact solutions. Using these elements, we look for the conditions such that the metric functions of the generalized scalar tensor theory tend toward exponential or power law solutions in the Einstein frame.
\section{Field equations} \label{s1}
In the Einstein frame, the metric can be written as: 
\begin{equation}\label{1}
ds^2=-(\bar{N}^2-\bar{N}_i \bar{N}^i )d\bar{\Omega}^2 + 2\bar{N}_i d\bar{\Omega}\omega^i + R_0 ^2 e^{-2 \bar{\Omega}}e^{2\beta_{ij}}\omega^i \omega^j
\end{equation}
\begin{equation}\label{1}
ds^2=-(\bar{N}^2-\bar{N}_i \bar{N}^i )d\bar{\Omega}^2 + 2\bar{N}_i d\bar{\Omega}\omega^i + R_0 ^2 e^{-2 \bar{\Omega}}e^{2\beta_{ij}}\omega^i \omega^j
\end{equation}
the $\omega^i$ being the 1-forms of the Bianchi type I model. The barred quantities are those of the Einstein frame. $\bar{N}$ and $\bar{N}_i$ are respectively the lapse and shift functions. The relation between the metric functions of the Einstein and Brans-Dicke frames is:
\begin{equation} \label{1a}
g_{ij}=\bar{g}_{ij}\phi^{-1}
\end{equation}
With $(i,j)=0,1,2,3$. Hence, in the Brans-Dicke frame, a potential $U$ of the Einstein frame can be written as:
\begin{equation} \label{1b}
U_{BD}=U\phi^2
\end{equation}
The Lagrangian of the generalized scalar tensor theory with a potential is given by:
\begin{equation} \label{2}
S=(16\pi)^{-1}\int \left[\bar{R}-(3/2+\omega(\phi))\phi^{,\mu}\phi_{,\mu}\phi^2 -U(\phi)\right]\sqrt{-\bar{g}}d^4 x
\end{equation}
where $\phi$ is a positive scalar field, $\omega(\phi)$ is the coupling function, and $U(\phi)$ is the potential. As the Universe is homogeneous, the scalar field depends on time variable only. We use the method employed in \cite{MatRyaTot73}\cite{Nar72} to find the ADM Hamiltonian. The ADM form of the action is written as:
\begin{equation} \label{3}
S=(16\pi)^{-1}\int (\pi^{ij}\frac{\partial \bar{g}_{ij}}{\partial \bar{t}}+\pi^{\phi}\frac{\partial \phi}{\partial \bar{t}}-\bar{N}C^0 -\bar{N}_i C^i )d^4 x
\end{equation}
the $\pi^{ij}$ and $\pi^\phi$ are, respectively, the conjugate momentum of the metric functions $\bar{g}_{ij}$ and the scalar field, $\bar{N}$ and $\bar{N}_i$ play the role of Lagrange multipliers. The quantities $C^0$ and $C^i$ are, respectively, the super-Hamiltonian and the supermomentum defined by: 
\begin{equation} \label{4}
C^0 =-\sqrt{^{(3)}\bar{g}}^{(3)}\bar{R}-\frac{1}{\sqrt{^{(3)}\bar{g}}}(\frac{1}{2}(\pi^k _k )^2 -\pi^{ij}\pi_{ij})+\frac{1}{\sqrt{^{(3)}\bar{g}}}\frac{\pi_\phi ^2 \phi^2 }{6+4\omega}+\sqrt{^{(3)}\bar{g}}U(\phi)
\end{equation}
\begin{equation} \label{5}
C^i =\pi^{ij}_{\mid j}
\end{equation}
the "$^{(3)}$" hold for the quantities calculated on the 3-space and the "$\mid$" for the covariant derivative in the 3-space. By varying the action with respect to $\bar{N}$ and $\bar{N}^i$ we find the two constraints $C^0=0$ and $C^i=0$. Then, by using them and the form of the metric functions, $\bar{g}_{ij}=R_0 ^2 e^{-2\bar{\Omega}}e^{2\beta_{ij}}$ with $(i,j)=1,2,3$, and after taking the surface integral $\int \omega^1 \wedge \omega^2 \wedge \omega^3$ equal to $(4\pi)^2 $ \footnote{This value is valuable for Bianchi type I and IX models.}, the action (\ref{3}) becomes:
\begin{equation} \label{6}
S=2\pi\int\pi^i _k d\beta_{ik}-\pi^k _k d\bar{\Omega}+1/2\pi_\phi d\phi
\end{equation}
The final form of the action is obtained by defining the traceless diagonal matrix $\beta_{ij}$ and $p_{ij}$ by following the procedure introducing by Misner \cite{Mis62}. We define:
\begin{equation} \label{6a}
p^i _k =2\pi\pi^i _k -\frac{2}{3}\pi\delta^i _k \pi^l _l
\end{equation}
and parameterise:
\begin{equation} \label{7}
6p_{ij}=diag(p_+ + \sqrt{3}p_- , p_+ - \sqrt{3}p_- , -2p_+ )
\end{equation}
\begin{equation} \label{8}
\beta_{ij}=diag(\beta_+ + \sqrt{3}\beta_- , \beta_+ - \sqrt{3}\beta_- , -2\beta_+ )
\end{equation}
Moreover, on the hypersurface of constant time, $\pi^{ij}_{\mid j}=0$ for the Bianchi type I and IX without rotation. Using the expression (\ref{6a})-(\ref{8}), the action (\ref{6}) can be written as:
\begin{equation} \label{9}
S=\int p_+ d\beta_+ + p_- d\beta_- + p_\phi d\phi - Hd\Omega
\end{equation}
with $p_{\phi}=\pi\pi_\phi$ and $H=2\pi\pi^k _k$. We can obtain the expression of the quantity $H$, that is $\pi^k _k$ from the constraint $C^0=0$. Then, we find for $H$:
\begin{equation} \label{11}
H^2 =p_+ ^2 +p_- ^2 +12\frac{p_\phi ^2 \phi^2}{3+2\omega}+36\pi^2 R_0 ^4 e^{-4\bar{\Omega}}(V-1)+24\pi^2 R_0 ^6 e^{-6\bar{\Omega}}U
\end{equation}
The potential $V(\beta_+ ,\beta_- )$ depends on the Bianchi model. For the Bianchi type I model, $V=1$. Finally the field equations for the generalized scalar tensor theory are Hamilton's equations for the Hamiltonian $H$:
\begin{equation} \label{12}
H^2 = p_+ ^2 +p_- ^2 +12\frac{p_\phi ^2 \phi^2}{3+2\omega}+24\pi^2 R_0 ^6 e^{-6\bar{\Omega}}U
\end{equation}
\begin{equation} \label{13}
\dot{\beta}_ \pm = \frac{\partial H}{\partial p_ \pm}=\frac{p_\pm}{H}
\end{equation}
\begin{equation} \label{14}
\dot{\phi}=\frac{\partial H}{\partial p_\phi}=\frac{12\phi^2 p_\phi }{(3+2\omega)H}
\end{equation}
\begin{equation} \label{15}
\dot{p}_\pm=-\frac{\partial H}{\partial \beta_ \pm}=0
\end{equation}
\begin{equation} \label{16}
\dot{p}_\phi=-\frac{\partial H}{\partial \phi}=-12\frac{\phi p_\phi ^2}{(3+2\omega)H}+12\frac{\omega_\phi \phi^2 p_\phi ^2 }{(3+2\omega)^2 H}-12\pi^2 R_0 ^6 \frac{e^{-6\bar{\Omega}}U_\phi }{H}
\end{equation}
\begin{equation} \label{17}
\dot{H}=\frac{dH}{d\bar{\Omega}}=\frac{\partial H}{\partial \bar{\Omega}}=-72\pi^2 R_0 ^6 \frac{e^{-6\bar{\Omega}}U}{H}
\end{equation}
where a dot denotes a derivative with respect to $\bar{\Omega}$. Moreover we will choose $\bar{N}^i=0$ and we express $\bar{N}$ by writing that $\partial \sqrt{\bar{g}}/\partial \bar{\Omega}=-1/2\pi^k_k\bar{N}$ (see \cite{Nar72}, p1830 for a detailed calculus). Hence, we find:
\begin{equation} \label{18}
\bar{N}=\frac{12\pi R_0 ^3 e^{-3\bar{\Omega}}}{H}
\end{equation}
Equation (\ref{15}) shows that the conjugate momemta $p_\pm$ are constants. Then, from the equations (\ref{13}), we deduce that $\beta_+-(p_+p_-^{-1})\beta_-$ is constant and the Universe point moves on a straight line in the $(\beta_+,\beta_-)$ plane. Since we have $d\bar{t}=-\bar{N}d\bar{\Omega}$ \footnote{We choose $d\bar{t}=-\bar{N}d\bar{\Omega}$ as in \cite{MatRyaTot73} but $d\bar{t}=\bar{N}d\bar{\Omega}$ is also a valid choice and would not change our results in $t$ or $\bar{t}$ times.}, equation (\ref{18}) shows that when the Hamiltonian has a constant sign, $\bar{\Omega}$ is a monotonous function of $\bar{t}$, decreasing if $H>0$ and increasing otherwise.
\section{Dynamical study of the metric functions in the proper time of the Einstein frame} \label{s2}
In this section we analyse the dynamics of the metric functions in the Einstein frame. They can be written:
\begin{equation} \label{18a}
\bar{g}_{ij}=R_0 ^2 e^{-2\bar{\Omega}+2\beta_{ij}}
\end{equation}
With $(i,j)=1,2,3$. Using $d\bar{t}=-\bar{N}d\bar{\Omega}$, we obtain:
\begin{equation} \label{20}
\frac{d\bar{g_{ij}}}{d\bar{t}}=2R_0^2 (\frac{d\beta_{ij}}{d\bar{t}}-\frac{d\bar{\Omega}}{d\bar{t}})e^{-2\bar{\Omega}+2\beta_{ij}}= 2R_0 ^2 e^{-2\bar{\Omega}+2\beta_{ij}}\frac{H-p_{ij}}{H\bar{N}}
\end{equation}
the product $H\bar{N}$ being positive. We are interested in the sign of the quantity (\ref{20}) which depends on the sign of $H-p_{ij}$. For sake of simplicity, we will consider a potential with a constant sign. We will see later how to extend our results to the case where the sign of the potential varies. Then, the equation (\ref{17}) shows that the sign of $H\dot{H}$ is constant and so for $H$ and $\dot{H}$. $H$ is a monotonic function of time and $p_{ij}$ is a constant, which means that equation (\ref{20}) can only have one zero. So if there is an extremum for the metric function when the potential is of constant sign, it is unique. $\bar{\Omega}$ is also a monotonic function of $\bar{t}$.

If $H_{ini}$ and $H_{fin}$ are the two values of the Hamiltonian at the extremities of the $\bar{t}$ proper time interval, H will evolve monotonically from $H_{ini}$ to $H_{fin}$. The first derivative (\ref{20}) of the metric function in the Einstein frame will vanish if the three conditions $C_1$, $C_2$ and $C_3$ are true:
\begin{itemize}
\item $C_1$: H and $p_{ij}$ have the same sign
\item $C_2$ and $C_3$: $p_{ij}$ belongs to the interval defined by $H_{ini}$ and $H_{fin}$
\end{itemize}
Hence, we have to consider the following four cases for which we give the variation of the metric function depending on the $\bar{t}$ time:
\newline
\newline
\emph{Case 1a: $U<0$, $\dot{H}$ and $H>0$}
\newline
We recall we have $d\bar{t}=-Nd\bar{\Omega}$. Hence taking into account (\ref{18}), when $\bar{\Omega}$ increases, $\bar{t}$ decreases. The Hamiltonian is a decreasing function of $\bar{t}$. The three conditions $C_i$ become:
\begin{itemize}
\item $C_1$: $p_{ij}>0$
\item $C_2$: $H_{fin}-p_{ij}$<0
\item $C_3$: $H_{ini}-p_{ij}$>0
\end{itemize}
Whatever case we consider, if $C_2$ or $C_3$ are false, respectively $C_3$ or $C_2$ are true. In addition, in the present case, if $C_3$ is false, $C_1$ is true. 

If the three conditions are true, the metric function has a maximum in the proper time of the Einstein frame since the Hamiltonian will be equal to $p_{ij}$ for a value of $\bar{\Omega}$.

If $C_1$ is wrong, it is increasing since then $H-p_{ij}$ has always the sign of $H$.

If $C_2$ is wrong, the Hamiltonian is always larger than $p_{ij}$, and the metric function is again increasing for the $\bar{t}$ time.

If $C_3$ is wrong, $C_1$ is true, and the metric function decreases since the Hamiltonian is always smaller than $p_{ij}$.
\newline
\newline
The same reasoning will hold for the other cases.
\newline
\newline
\emph{Case 1b: $U<0$, $\dot{H}$ and $H<0$}
\newline
When $\bar{\Omega}$ increases, $\bar{t}$ is increasing. The Hamiltonian is a negative and decreasing functions of these times coordinates. When $C_2$ is wrong, $C_1$ is true. The three conditions can be written as:
\begin{itemize}
\item $C_1$: $p_{ij}<0$
\item $C_2$: $H_{fin}-p_{ij}$<0
\item $C_3$: $H_{ini}-p_{ij}$>0
\end{itemize}

If they are all true, the metric function has a maximum.

If $C_1$ or $C_3$ is false, it is decreasing.

If $C_1$ is true, $C_2$ is false and it is increasing.
\newline
\newline
\emph{Case 2a: $U>0$, $\dot{H}<0$ and $H>$0}
\newline
When $\bar{\Omega}$ increases, $\bar{t}$ decreases. The Hamiltonian is a positive and decreasing function of $\bar{\Omega}$ and then an increasing function of $\bar{t}$. When $C_2$ is wrong, $C_1$ is true. The three conditions can be written as:
\begin{itemize}
\item $C_1$: $p_{ij}>0$
\item $C_2$: $H_{fin}-p_{ij}$>0
\item $C_3$: $H_{ini}-p_{ij}$<0
\end{itemize}

If they are all true, the metric function has a minimum.

If $C_1$ or $C_3$ is false, it is increasing.

If $C_2$ is false, $C_1$ is true, and the metric function is decreasing.
\newline
\newline
\emph{Case 2b: $U>0$, $\dot{H}>0$ and $H<$0}
\newline
When $\bar{\Omega}$ increases, $\bar{t}$ increases. The Hamiltonian is a negative and increasing function of the two time coordinates. When $C_3$  is false, $C_1$ is true. We obtain for the three conditions:
\begin{itemize}
\item $C_1$: $p_{ij}<0$
\item $C_2$: $H_{fin}-p_{ij}$>0
\item $C_3$: $H_{ini}-p_{ij}$<0
\end{itemize}

If the three conditions are true, the metric function has a minimum.

If $C_1$ or $C_2$ is false, it is decreasing.

If $C_1$ is true, $C_3$ is false, the metric function is increasing.
\newline
\newline
All these results are summarised in table 1.
\newline
\begin{table}[p]
\caption{Dynamical behaviour of a metric function in the proper time of the Einstein frame depending on the signs of the potential, the Hamiltonian and its initial and final values.}
\begin{indented}
\item[]\begin{tabular}{@{}lllll}
\br
 & $H$, $\dot{H}>0$, $U<0$ & $H$, $\dot{H}$, $U<0$ & $H$, $U>0$, $\dot{H}<0$ & $\dot{H}$, $U>0$, $H<0$\\
\mr
$C_1$, $C_2$, $C_3$: true & Maximum & Maximum & Minimum & Minimum 		\\
$C_1$: false 		& Increasing & Decreasing & Increasing & Decreasing \\
$C_2$: false 		& Increasing & 		& 		& Decreasing \\
$C_3$: false 		&		 & Decreasing & Increasing & 		  \\
$C_2$: false, $C_1$: true & & Increasing & Decreasing & 				  \\
$C_3$: false, $C_1$: true & Decreasing & 		&		 & Increasing \\
\br
\end{tabular}
\end{indented}
\end{table}
\newline
Before analysing this table, lets note that we will use the expression "Big-Bang singularity" to denote the fact that the three metric functions decrease toward zero. In addition the expression "pancake singularity" or "cigar singularity" apply, respectively, to the cases where one or two metric functions decrease toward zero.

From the table 1, we obtain the following results in the Einstein frame. We deduce that a metric function could have a maximum (minimum) only in the presence of a negative (positive) potential. Moreover, all the conjugate momentum $p_{ij}$ can not have the same sign and then the condition $C_1$ can not be true for all the metric functions. We deduce that, when the Hamiltonian is positive, the three metric functions can be increasing together at late times, but not decreasing. All types of singularity, Big-Bang type, pancake type or cigar type are possible at early time. When the Hamiltonian is negative, the three metric functions can be decreasing together at late time but not increasing. The singularity if it exists will only be of pancake or cigar type at early time. We have already written that as long as the potential has a constant sign, the metric function can have one and only one extremum. This is also the case when we consider flat or open FLRW models with trace-free matter, $\phi$ finite and $\omega_\phi>0$ as shown in \cite{BarPar97}. In this paper it is also proved that flat FLRW models can only contain a single minimum whereas here, a single maximum is also allowed for negative potential. 

Lastly, it is easy to calculate that $d\beta_\pm/d\bar{t} \propto e^{3\bar{\Omega}}$. This means that the Universe will isotropize, that is $\bar{g}_{ij}/(d\bar{g}_{ij}/dt)$ tends toward the same function whatever i and j, only when $\bar{\Omega}\rightarrow -\infty$. This value will correspond to late (early) times for $\bar{t}$ if the Hamiltonian is positive (negative).

When the sign of the potential varies, the table is always true but $H_{ini}$ and $H_{fin}$ define the different intervals of values of the Hamiltonian for which the sign of the potential is constant. Hence, if asymptotically the sign of the potential is constant, one can always use the previous results.
\section{Necessary and sufficient conditions to obtain an isotropic Universe in expansion at late time in the Brans-Dicke frame.} \label{s3}
In this section we look for isotropisation and expansion of the metric functions at late time in the Brans-Dicke frame. Let $t_0$ be the maximum value (finite or not) of the t-time coordinate, that is the value of $t$ at late time. We suppose that the physical conditions in $t_0$ are the same as those of today. Hence the scalar field will be such that $\omega>500$, $\omega_{\phi}\omega^{-3}\rightarrow 0$ and $U\rightarrow 0$ or very small. These conditions have been assumed to be necessary so that the relativistic values of the PPN parameters are respected. In a Universe without any matter field \cite{Nor68}\cite{Wag70} they can be written as:
\begin{equation} \label{20c}
\beta=1+O(\omega_\phi\omega^{-3})
\end{equation}
\begin{equation} \label{20b}
\gamma=1-(\omega+2)^{-1}
\end{equation}
Since in the generalized scalar tensor theory the inverse of the scalar field can be considered like the gravitational coupling function $G$, we assume that it tends toward a positive constant at late times. This is justified by measurements of the quantity $\dot{G}G^{-1}$. For a review of these experiments see \cite{BarPar97}. Hence, a relativistic limit shall be asymptotically recovered. 

A necessary and sufficient condition such that the Universe is isotropic at late time, that is $g_{ij}/(dg_{ij}/dt)$ tends toward the same function whatever i and j, will be:
\begin{equation} \label{20a}
d\beta_\pm/dt\propto e^{3\bar{\Omega}}\sqrt{\phi}\rightarrow 0
\end{equation}
that is $\beta_\pm$ tend toward a constant. Then, the three metric functions are proportional to the function $e^{-2\bar{\Omega}}$ in the Einstein frame or $e^{-2\bar{\Omega}}\phi^{-1}$ in the Brans-Dicke frame. If we want that the Universe be in expansion at late times, this last function have to be increasing in the Brans-Dicke frame when $t\rightarrow t_0$. We write the derivative of this function with respect to $\bar{\Omega}$:
\begin{equation} \label{21}
(e^{-2\bar{\Omega}}\phi^{-1})^{.}=-e^{-2\bar{\Omega}}\phi^{-1}(\frac{\dot{\phi}}{\phi}+2)
\end{equation}
There are two ways so that it can be increasing in the t-time. Firstly, we suppose that $t_0$ coincides with an infinite value of $\bar{\Omega}$.
\newline
\newline
If $\dot{\phi}\phi^{-1}>-2$ when $t\rightarrow t_0$, $e^{-2\bar{\Omega}}\phi^{-1}$ is a decreasing function of $\bar{\Omega}$ and it will be an increasing function of $t$ if the Hamiltonian is positive. This means that $t\rightarrow t_0$ coincides with $\bar{\Omega}\rightarrow -\infty$. We note that for a decreasing scalar field on the $t$ time there are no additional conditions coming from the fact that $\dot{\phi}\phi^{-1}>-2$ whereas an increasing one have to respect $\dot{\phi}\phi^{-1}\in\left[-2,0\right]$ when $t\rightarrow t_0$ \footnote{We would have the inverse situation if we had considered a negative scalar field.}. Hence, in a Universe undergoing expansion at late times in the Brans-Dicke frame, increasing scalar field $\phi(t)$ implies fine-tuning.

As the scalar field tends toward a constant and $\bar{\Omega}\rightarrow -\infty$, equation (\ref{20a}) shows that the Universe isotropizes in a natural way, that is without any other condition.
\newline
\newline
If now $\dot{\phi}\phi^{-1}<-2$ when $t\rightarrow t_0$, $e^{-2\bar{\Omega}}\phi^{-1}$ is an increasing function of $\bar{\Omega}$ and it would be an increasing function of $t$ if the Hamiltonian were negative. This means that $t\rightarrow t_0$ coincides with $\bar{\Omega}\rightarrow +\infty$. The scalar field is always a decreasing function of $t$. 

The relation (\ref{20a}) shows that the Universe will isotropize at $t_0$ if then $\phi<e^{-6\bar{\Omega}}$.
\newline
\newline
Secondly, if we consider that the $t_0$ time coincides with a finite value of $\bar{\Omega}$, we can write the same conditions so that the function $e^{-2\bar{\Omega}}\phi^{-1}$ is an increasing function of $t$ at late times depending on the sign of $\dot{\phi}\phi^{-1}+2$, but to obtain an isotropic Universe the scalar field has to vanish in $t_0$ since from (\ref{20a}) we see that $d\beta_\pm/dt$ is now proportional to $\phi^{1/2}$.
\newline
\newline
Another fact to take into account to obtain a realistic Universe at late $t$ time is the recently observed accelerated dynamics of the Universe which implies a positive cosmological constant. So that the potential, in Einstein or Brans-Dicke frames, is positive at late time, we deduce from (\ref{17}) that when the Hamiltonian is positive (negative), it is a decreasing (increasing) function of $\bar{\Omega}$ and hence an increasing function of $t$.
\newline
\newline
Finally we summarise these results in table 2. 
\newline
\newline
From the above, we deduce the following results in the Brans-Dicke frame. When $\bar{\Omega}$ diverges at late t-time, the Universe of the Bianchi type I model, in the generalized scalar tensor theory and in the Brans-Dicke frame, with a positive potential will isotropize and be in expansion if $\dot{\phi}\phi^{-1}>-2$ and the Hamiltonian is a positive and increasing function of the t-time. If $\dot{\phi}\phi^{-1}<-2$, the Hamiltonian have to be a negative and increasing function of the t-time and the scalar field has to be less than $e^{-6\bar{\Omega}}$. If $\bar{\Omega}$ tends toward a constant at late t-time, we need $\dot{\phi}\phi^{-1}>-2$  ($\dot{\phi}\phi^{-1}<-2$ ), a positive and increasing (negative and increasing) Hamiltonian in the t-time and a vanishing scalar field. Let us note, that a Universe able to isotropize at both late and early times can exist.

\begin{table}[p]
\caption{Conditions for the Universe to be isotropic, in expansion and with a positive potential at late time in the Brans-Dicke frame.}
\begin{indented}
\item[]\begin{tabular}{@{}lllll}
\br
 & Expansion in $t_0$ & Isotropisation in $t_0$ & $U_{BD}>0$ in $t_0$ \\
\mr
$\bar{\Omega}$ diverges & $\dot{\phi}/\phi>-2$, $H>0$: $\bar{\Omega}\rightarrow -\infty$ & Yes & $dH/dt>0$ or $dH/d\bar{\Omega}<0$ \\
 & $\dot{\phi}/\phi<-2$, $H<0$: $\bar{\Omega}\rightarrow +\infty$ & Yes if $\phi<e^{-6\bar{\Omega}}$ & $dH/dt>0$ or $dH/d\bar{\Omega}>0$ \\
$\bar{\Omega}\rightarrow cte$ & $\dot{\phi}/\phi>-2$, $H>0$ & Yes if $\phi\rightarrow 0$ & $dH/dt>0$ or $dH/d\bar{\Omega}<0$ \\
 & $\dot{\phi}/\phi<-2$, $H<0$ & Yes if $\phi\rightarrow 0$ & $dH/dt>0$ or $dH/d\bar{\Omega}>0$ \\
\br
\end{tabular}
\end{indented}
\end{table}
\emph{\underline{Remark}: }All the results of table 2 are expressed in the $\bar{\Omega}$-time except the sign of H and $dH/dt$. By defining the 3-Volume V in the Brans-Dicke time by $V=e^{-3\bar{\Omega}}\phi^{-3/2}=det\sqrt{^{(3)}g}\phi^{-3/2}$ one can also write the condition on the sign of $\dot{\phi}\phi^{-1}+2$ with physical quantities of the Brans-Dicke time. By writing that $\dot{\phi}=\frac{d\phi}{dt}\frac{dt}{d\bar{\Omega}}$, this expression becomes:
\begin{equation} \label{22a}
\frac{\dot{\phi}}{\phi}+2=\frac{d\phi}{dt}\phi^{-1}\left[ln(V^{-1/3}\phi^{1/2})\right]^{-1}
\end{equation}
\section{Discussions} \label{s4}
Numerous works have been devoted to the problem of the physical frame between the Brans-Dicke or Einstein frame \cite{RaiZhu98, Sok95, QuiBonCar99}. In the Brans-Dicke frame, the scalar field is related to the gravitational coupling function and is non-minimally coupled to the gravitational field. In the Einstein frame, the scalar field is associated with the rest mass of the particles and is minimally coupled to the gravitational field. Lets examine the optimal conditions in each frame to obtain asymptotically an isotropic expanding Universe with a positive potential.\\
In the Einstein frame, we need a positive Hamiltonian so that the three metric functions are increasing at late times. The potential will be positive if $H$ is a decreasing function of $\bar{\Omega}$ and then an increasing one of $\bar{t}$. In these conditions, no more than two metric functions can have one and only one minimum. All types of singularity are possible at early time. Since the Universe isotropizes if $\bar{\Omega}\rightarrow -\infty$ and as $H>0$, it will arise at late times.\\
In the Brans-Dicke frame, an expanding and isotropic Universe with a positive potential at late times can be realized in four different ways described in table 2. However, only one of them does not need the scalar field to vanish asymptotically. It is such that the Hamiltonian has the same features as in the Einstein frame with $\bar{\Omega}\rightarrow -\infty$ and $\dot{\phi}/\phi>-2$. It is important to avoid the scalar field vanishing because it would mean that the gravitational constant is asymptotically infinite. However its current observed value seems to be small and constant.\\
Hence, this work does not allow us to argue in favour of one of the frame since the Hamiltonian and time $\bar{\Omega}$ have the same features in both frames, corresponding to the dynamics and properties of the Universe we want to obtain at late time, that is isotropy, expansion and positive potential. This is not a surprise because, when we have studied the late time behaviour of the metric functions in the Brans-Dicke frame, we have assumed that asymptotically the scalar field tended toward a constant. Thus, at late time, the two frames become similar.
\newline
\newline
Before carrying on with this discussion, it is useful to know how to find exact solutions from the system of equations (\ref{12})-(\ref{18}) by making use of our previous results. In the generalized scalar tensor theory with a potential, two functions can be chosen arbitrarily to completely define the theory. The method we will use is the following:\\
We choose $U(\bar{\Omega})$ or $H(\bar{\Omega})$ and we determine respectively the Hamiltonian or the potential with (\ref{17}) and then the functions $\beta_\pm$ with (\ref{13}). Then, we choose $\omega(\bar{\Omega})$ or $\phi(\bar{\Omega})$  and with (\ref{12}), we obtain respectively the scalar field or the coupling function. With the help of (\ref{18}) and (\ref{1a}), we find $\bar{\Omega}(\bar{t})$ and $\bar{t}(t)$ and then the expressions of each quantity in the proper time of each frame. Since we have determined what are the characteristics of $H$, $\phi$ and $\bar{\Omega}$ to obtain physically interesting late time behaviour, it is easy to obtain as many exact solutions as we want with isotropic expanding behaviour and positive potential. \\
Other methods such as dynamical ones could be used to study the equations (\ref{12})-(\ref{18}) since they constitute a  system of first order differential equations. However our goal is to find conditions to obtain an asymptotically isotropic expanding Universe with a positive potential and here such a method is not necessary. Application of dynamical methods to the system of equations (\ref{12})-(\ref{18}) will be the subject of future works. Some more powerful methods to derive exact solutions from Hamiltonian formalism have been developed in \cite{UggJanRos95}. They rely on symmetries such as Killing tensor symmetries. However, it is difficult to predict the late time behaviour of the solutions thus obtained and, if they are very efficient when a perfect fluid is present, it is different if we consider any potential. The method explained above has the advantage of predicting the late time behaviour of the solution once the two unknown functions fixed thanks to the results of the previous sections.
We will use it to examine two important asymptotical behaviours in the Einstein frame for the metric functions: exponential and power-law behaviours. We have chosen to study them in the Einstein frame rather than in the Brans-Dicke frame since we will be able to compare our results with those obtained in General relativity with a scalar field.
\newline
\newline 
Firstly, we examine the exponential behaviour for the metric functions. Then, we shall try to recover the "No Hair Theorem" for the Bianchi type I model so that we test our results. Wald \cite{Wal83} has shown that, in the case of General Relativity with a scalar field and a cosmological constant, all the Bianchi models (except contracting Bianchi type IX) initially in expansion approach the isotropic De Sitter solution. If we consider the Generalized scalar tensor theory in the Einstein frame, we obtain a positive cosmological constant by choosing $H=\Lambda e^{-3\bar{\Omega}}$ with $\Lambda>0$. Then, the metric functions are:
\begin{equation}
\bar{g}_{ij}=e^{\Lambda(6\pi R_0^3)^{-1}(\bar{t}-\bar{t}_0)+2p_{ij}(3\Lambda)^{-1}e^{-\Lambda(4\pi R_0^3)^{-1}(\bar{t}-\bar{t}_0)}+2\beta_{ij0}} \nonumber
\end{equation}
$\beta_{ij0}$, $\bar{t}_0$ and $p_{ij}$ being some constants. $\bar{\Omega}$ varies from $+\infty$ to $-\infty$ and $\bar{t}$ respectively from $-\infty$ to $+\infty$. At late times, whatever the coupling function such that $\phi(\bar{\Omega})$ is defined for $\bar{\Omega}\rightarrow -\infty$, the Universe will isotropize and approach a De Sitter model in accordance with Wald. The properties of the Hamiltonian and the time $\bar{\Omega}$ correspond to those we have defined for this type of behaviour in section \ref{s2}. At early time, when $\bar{\Omega}\rightarrow +\infty$, the $\beta_\pm$ functions dominate the dynamical behaviour, and the singularity will be of cigar or pancake type. If we choose $\Lambda<0$ the behaviour of the late and early times are inverted.\\
Now, we make the opposite reasoning. We suppose that at late time, the 3-volume has an exponential behaviour. We want to know whether the Universe will isotropize and whether the potential and the coupling constant respect the solar system tests at late time. As we know the form of the 3-volume asymptotically, we can determine that of the Hamiltonian $H(\bar{\Omega})$ from $d\bar{t}=-\bar{N}d\bar{\Omega}$. With this expression, we can check that for a general asymptotical form $1/f(\bar{t})$ of the 3-volume, $H^{-1}$ will be equal asymptotically to $(-12\pi R_0^3)^{-1}(\left[A(\bar{\Omega})G(\bar{\Omega})\right]^.+\dot{B}(\bar{\Omega}))$, where $A$ and $B$ are any function such that $A\rightarrow 1$, $B\rightarrow 0$ when $\bar{\Omega}\rightarrow -\infty$, $G=F(f^{-1}(e^{3\bar{\Omega}}))$ and $F=\int f(\bar{t})d\bar{t}$. Here, $1/f(\bar{t})=e^{3\bar{t}}$ and $G=-1/3e^{3\bar{\Omega}}$. We will choose a class of Hamiltonian functions such that $H$ and $\dot{H}$ do not oscillate at late times that is the first and second derivatives of $A$ and $B$ vanish asymptotically (however our results will be the same for types of functions such $\cos(\bar{\Omega}^{-1})$ and $\sin(\bar{\Omega}^{-1})$ corresponding respectively to $A$ and $B$ with damped oscillations and which have the same asymptotic characteristics described above). Hence, the Hamiltonian tends toward $e^{-3\bar{\Omega}}$. This form excludes any oscillating potential at late time. From (\ref{17}), we deduce that $U\rightarrow C^2$, where $C$ is a constant. It follows from the results of section \ref{s2}, that if the scalar field is also defined in $\bar{\Omega}\rightarrow -\infty$, the Universe isotropize at late time, corresponding to this last value of $\bar{\Omega}$. Then it tends toward a De Sitter behaviour and the potential toward a positive constant. This generalizes the result of Wald for Bianchi type I model to any potential that is asymptotically constant and does not oscillate. Using (\ref{12}) and (\ref{14}), we show that asymptotically, $3+2\omega\propto \phi^2\dot{\phi}^2$ and $\omega_\phi\omega^{-3}\propto \dot{\phi}^4\phi^{-5}-\ddot{\phi}\dot{\phi}^2\phi^{-4}$. If $\phi$ tends toward a non-vanishing constant, then the coupling function and $\omega_\phi\omega^{-3}$ respectively diverge and vanishes asymptotically. This limit for the scalar field is the most interesting one since it is proportional to the inverse of the gravitational function. This leads to the fact that any $\omega$ satisfies the solar system tests for such a limit reached in $\bar{\Omega}\rightarrow -\infty$. It will also be the case for the potential if the constant $C^2$ is sufficiently small. Other limits for $\phi$ could be envisaged and not be in contradiction with the previous quoted tests or isotropisation in Einstein frame. Above all $\phi\rightarrow \infty$ which leads to an asymptotically vanishing gravitational constant. We will present some examples in the next paragraph.\\
Another interesting behaviour for the 3-volume is a power law one since more often we search for theories which tend toward General relativity at late times and since this last one, in isotropic and flat cases, has most of time power law solutions. Let us have a look at what happens when the 3-volume of the Universe tends toward a power law form of $\bar{t}$ at late time, that is $e^{-3\bar{\Omega}}\propto \bar{t}^{3m}$. We proceed in the same way as previously. We deduce that asymptotically the Hamiltonian tends toward $e^{l\bar{\Omega}}$ with $l=m^{-1}-3$. To obtain a positive potential we shall have $l<0$, that is $m\not\in\left[0,1/3\right]$. Then, if the scalar field is defined in $\bar{\Omega}\rightarrow -\infty$, the Universe isotropizes and the metric functions tend toward a power law $\bar{t}^{2m}$. The Universe is in expansion if $m>0$ that is $l>-3$ and undergoes inflation if $m>1$, that is $l\in\left[-3,-2\right]$. In what it follows, we assume that $l$ belongs to $\left[-3,0\right]$. From (\ref{17}) we deduce that the potential is proportional to $e^{(2l+6)\bar{\Omega}}$. At late times, it vanishes in agreement with solar system tests, whatever $l$. Concerning the coupling function $\omega$, we have the same limits as above as long as $l<0$ and we can write the same things. However, here we shall also use the fact that $\dot{\phi}=\dot{U}U_\phi^{-1}$ and $\ddot{\phi}=\ddot{U}U_\phi^{-1}-\dot{U}^2U_{\phi\phi}U_\phi^{-3}$. However, at late time $\dot{U}=(2l+6)U$ and $\ddot{U}=(2l+6)^2U$. Thus, asymptotically $3+2\omega\propto\phi^2U_\phi^2U^{-2}$ and $\omega_\phi\omega^{-3}\propto U^3\left[UU_\phi+\phi(UU_{\phi\phi}-U_\phi^2)\right]U_\phi^{-5}\phi^{-5}$. So, for any given form of $U(\phi)$, we can determine whether the solar system tests will be recovered as the Universe isotropize in $\bar{\Omega}\rightarrow -\infty$. As an application, we examine two typical forms for the potential: $U=e^{k\phi}$ and $U=\phi^k$. For the first form, the scalar field shall tend toward $(2l+6)k^{-1}\bar{\Omega}$ and for the second one toward $e^{(2l+6)k^{-1}\bar{\Omega}}$. Both limits are defined for $\bar{\Omega}\rightarrow -\infty$. Thus, the forms we choose for the potential are compatible with a scalar field defined in $\bar{\Omega}\rightarrow -\infty$. Firstly, we examine $U=e^{k\phi}$ with $k>0$. Such potentials are well motivated, especially from string theory. They are also used to generate scaling solutions for which the energy density of the scalar field mimics the equation of state of a barotropic fluid \cite{Wet88} although they are not necessary well adapted \cite{PieDem99} to this type of problem. Asymptotically, the scalar field diverges and the potential vanishes. Hence, $\omega$ and $\omega_\phi\omega^3$ respectively diverges and vanishes for $\bar{\Omega}\rightarrow -\infty$. A theory with the same types of potential and behaviour at late times for the Universe has been studied in \cite{KitMae92, ColIbaHoo97}. However the coupling function was a constant and did not diverge at late times. Hence the corresponding Hamiltonian does not belong to the class we used in this work and will probably be oscillating at late times.  Secondly, we examine $U=\phi^k$ with $k<0$. Recently, this type of potential has been use to generate scaling solutions too \cite{HolWan99}. Again the scalar field diverges asymptotically and the potential vanishes. At late time $\omega$ becomes a constant and $\omega_\phi\omega^{-3}$ vanishes. Thus, for these types of potentials, the scalar field is defined in $-\infty$ where it diverges and $\omega$ and $U$ respect the solar system tests.\\
\newline
We conclude this discussion by summarising these results. We have shown that to obtain an isotropic expanding Universe at late times with a positive potential, we shall have $H>0$, $\dot{H}<0$ and $\bar{\Omega}\rightarrow -\infty$. This is necessary and sufficient for Einstein frame, sufficient and better for the Brans-Dicke frame since the gravitational constant does not diverge.\\
We have presented a method to obtain exact solution in the two frames. Then, considering the Einstein frame, we have recover Wald's theorem for Bianchi type I model.\\
The next results have been obtained by making the assumptions that the coupling function was such that the scalar field be defined in $\bar{\Omega}\rightarrow -\infty$ and that the Hamiltonian and thus the potential do not oscillate at late times.\\
Then, we have proved that when the 3-volume behaves asymptotically like an exponential, the Universe isotropizes toward a De Sitter model and the potential became asymptotically a constant. Moreover, if at late times the scalar field is a constant different from zero, which seems to be a physically reasonable assumption if we consider measurements of the gravitational constant, the values of $\omega$ and $\omega_\phi\omega^{-3}$ are in agreement with the solar system tests. Reciprocally, when a non-oscillating potential becomes a constant asymptotically, the Universe tends toward a De Sitter model whatever $\omega$ in accordance with our assumptions. This generalizes Wald's result for the Bianchi type I model and shows that the De Sitter model is an attractor for this class of potential.\\
When the 3-volume behaves asymptotically as a power law of $\bar{t}$, the Universe isotropizes and the metric functions tend toward $\bar{t}^{2m}$. The potential will be positive if $m>1/3$ and will vanish asymptotically. Thus, such a type of Universe solves the cosmological constant problem naturally. This enlightens the importance of power-law solutions in cosmology. If we assume that $\phi$ tends toward a constant, once more again, the coupling function respects the solar system tests. We can also express $\omega$ and $\omega_\phi\omega^{-3}$ asymptotically as some functions of $\phi$, the potential and its derivative with respect to the scalar field. Then, we have shown that for an exponential potential $e^{k\phi}$ with $k>0$, the coupling function and $\omega_\phi\omega^{-3}$ were in agreement with the solar system tests. $\omega$ can not be a constant since it diverges and thus, such theory will not tend toward a Brans-Dicke one. For a power law potential $\phi^k$ with $k<0$, the coupling function tends toward a constant and $\omega_\phi\omega$ vanishes. So, the theory can tend toward Brans-Dicke theory and this constant have to be larger than 500 so that the theory respects the solar system tests at late times. For these two types of potentials, the scalar field diverges. We have checked that its asymptotic form was defined in $\bar{\Omega}\rightarrow -\infty$.

\ack
To the memory of Jacques Demaret

\end{document}